\documentclass{JHEP3}

\usepackage{amsmath}
\input epsf
\usepackage{epsfig}
\usepackage{amssymb}
\usepackage{graphics}
\usepackage[active]{srcltx}
\usepackage{graphicx}

\newcommand{\bea}{\begin{eqnarray}}
\newcommand{\eea}{\end{eqnarray}}
\newcommand{\bean}{\begin{eqnarray*}}
\newcommand{\eean}{\end{eqnarray*}}
\newcommand{\nn}{\nonumber \\}

\renewcommand{\H}{\mathcal{H}}

\newcommand{\N}{\mathcal{N}}

\renewcommand{\d}{\delta}

\renewcommand{\d}{\partial}
\newcommand{\be}{\begin{eqnarray}}
\newcommand{\ee}{\end{eqnarray}}

\def\braket#1{\left\langle #1 \right\rangle}

\def\Tr{\mathop{\rm Tr}}

\def\d{{\rm d}}

\def\d{\partial}

\def\Label#1{\label{#1}%
  \smash{\hbox to0pt{\raise1ex\hbox{\tiny[#1]}\hss}}}

\title{Realize Emergent Gravity to Generic Situations}
\author{Yang An$^{\sharp}$,Peng Cheng$^{\sharp}$\\
~~~~\\
$^{\sharp}$Institute for Theoretical Physics, University of Amsterdam, Science Park 904,
1098 XH Amsterdam, The Netherlands\\
E-mail: {\tt anyangpeacefulocean@zju.edu.cn, p.cheng@uva.nl }\\ \\
}

\abstract{
We clarify the problem in which occasions 
can gravitational force be regarded emergent from thermodynamics, by proposing an entropic mechanism that can extract the entropic gradient existing in spacetime, due to the variation of the Casini-Bekenstein bound in specific quasi-static processes with the heat flux $\delta Q$ into the whole casual wedge. We explicitly formulate the derivation of inertial force as the emergent gravitational attraction from the Entanglement First Law. 

We find the saturation of the bound along with the vanishing relative entropy corresponds to the variation of minimal surface. To covariant meaning, it is the Bousso bound. Besides, this understanding is connected to recent Pennington’s work on Black Hole Information Paradox, suggesting a Page-Curve function origins from removing attraction by the external heat bath.

Our theory from entanglement now overcomes several criticism towards Verlinde's original entropic force proposal, and is able to co-exist with Susskind's Complexity Tendency. This entropic mechanism reproduces the Newton's Second Law in Rindler space and the gravitational force (together with derivation of the Einstein equation) beyond the near-horizon region, and can be adapted into AdS/CFT and other generic situations. 

}
\begin{document}

\section{Introduction}

Gravitational force is special, whose origin may be approached in a totally different way from other kinds of fundamental forces, which have been quantized and unified. Spacetime and gravity has been regarded as an emergent phenomenon from microscopic degrees of freedom of quantum field theory, 
an insight from the developments of string theory and loop quantum gravity, the two potential candidates of quantum gravity. Question is raised if the gravitational attraction reflects a fundamental tendency of information? 

From the AdS/CFT correspondence \cite{Maldacena:1997re}, early attempts \cite{Swingle:2009bg,VanRaamsdonk:2009ar,VanRaamsdonk:2010pw} show entanglement builds spacetime geometry, in the sense that the connection and continuity of spacetime geometry is closely related to the entanglement structure of QFT states. The idea of entanglement generating spacetime \cite{VanRaamsdonk:2010pw} then leads to the conjectures of $A=R_B$ \cite{Verlinde:2012cy} and then ER=EPR \cite{Maldacena:2013xja}. They were proposed to save the Principle of Equivalence against the Firewall Paradox in AMPS \cite{Almheiri:2012rt} argued from the monogamy of entanglement.
While, a general rule holds for any quantum system, the entanglement first law is then applied to gravity, and leads to breakthrough results, the derivation of Einstein Equation from AdS/CFT to linearized level \cite{Lashkari:2013koa,Faulkner:2013ica} as well as to non-linear level \cite{Faulkner:2017tkh}, and Jacobson's new derivation \cite{Jacobson:2015hqa}  of Einstein Equation based on Maximal Vacuum Entanglement Hypothesis. 
But since  those developments are based on vacuum entanglement, they are not equivalent to explain the tendency of gravitational attraction. One should apply this entanglement first law for perturbing excited states to reconsider old questions put up in Verlinde's emergent gravity theory.

One decade ago, Verlinde remarkably attached information meaning to gravitational attraction through the entropic force conjecture \cite{Verlinde:2010hp}.
 The basic idea of Verlinde's emergent gravity theory \cite{Verlinde:2010hp,Verlinde:2016toy} is that the gravitational force is possibly an entropic force $F=T\nabla S$ that usually occurs in macroscopic systems such as colloid and polymer molecules,  with the entropy gradient from variation of "holographic screen" existing generally in spacetime. In this way, the theory is in an attempt to explain the falling tendency of Newton's apples as entropy-increasing tendency of the thermodynamic second law.

However, this theory is rather controversial and under criticism. It requires either subtle improvement or modification, since the reason for the existence of the entropic gradient remains unclarified. Even, chances are that such entropic mechanism may not exist at all: it is possible to get $dW=dE$ with no entropy varying $dS=0$. Such querying and doubt about the ability of entropic mechanism to explain was put, for example, in \cite{Dai:2017qkz} and \cite{Susskind:2019ddc}. 
On the other hand, so far it couldn't explain the gravitational attraction in generic situations, for the original holographic screen approach fails to be generalized beyond near-horizon region, and it can't be applied to the sun nor planets. Through calculating the back-reaction to the geometry, the conjecture is tested only to hold in near-horizon regions in \cite{Bhattacharya:2018wfr}, for area variation of the "holographic screen" provides too much entropy.  
In all, how to interpret gravitational force from an entropic mechanism still remains a mystery.

The argument continues recently, after Susskind proposed a new alternative description in \cite{Susskind:2018tei} from Complexity Tendency, together with a query of the explanation ability of the entropic mechanism for the oscillating movement in pure AdS \cite{Susskind:2019ddc}. Whether the gravitational force can be interpreted from an entropic mechanism even becomes a question.

The resolution of the snags could be simple. To match the local gravitational force $F_\mu^g=-\frac{GMm}{r^2(1-\frac{{2GM}}{r})}\delta^r_\mu$ derived for Schwazchiled solution in GR (see textbook \cite{Carroll:2004,Wald:1984}), our former research \cite{An:2018hyt} happened to work out in a simple single-mode thermal harmonic oscillator model. It suggests the entropic change is exactly the variation of Casini-Bekenstein bound \cite{Casini:2008cr}.
This work was inspired from the observation that the process of a static observer lifting/lowing a box through a long string in Bekenstein's famous thought experiment for the Generalized Second Law \cite{Bekenstein:1974ax}, is indeed a quasi-static process, like the processes that a heat engine endures during the Carnot cycle, after considering thermalization of the box by the local Hawking temperature.
Therefore this non-unitary process changes the entanglement entropy within the global casual wedge, then causes heat flow $\delta Q=T\delta S$ into the exterior region of the black hole, through external influence by the long string.

What make a difference is to include the thermalization by the local Hawking temperature $T=T_H/V$ to replace the usual rule of the Unruh temperature in Entropic Gravity theories, to the box regarded as an excited state confined in the subsystem. Since Hawking/Unruh effect happens to different static observers related by the redshift factor $V$, the entropic gradient comes out along with the temperature gradient when the string slowly moves the box. Then to calculate inertial force, one should adopt the entanglement first law involving excited states' modular hamiltonian if we consider the entanglement entropy during this process.



In this paper, we will illustrate this entropic mechanism in more details and show it can be developed to explain gravitational attraction in generic situations. 

\paragraph{Main Results}

The primary problem to solve, is to find what causes the entropy variation, then we may be able to calculate how much should such variation be in general. More specifically, we should find which well-defined entropy is necessary for the gravitational force, and under which thermodynamic process.

Based on the positivity of relative entropy, Casini proved a more concise version of the Bekenstein bound \cite{Casini:2008cr} for any relativistic Quantum Filed Theory
\bea
\Delta S\leq \Delta \braket{K}\,.
\label{CBbound}
\eea
which is related to the modular Hamiltonian $K$ and the entanglement entropy. The proof is for the Rindler space of Minkowski spacetime, but also applied to eternal Schwarzschild black hole that has Hartle-Hawking states as its vacuum. Our main derivation in this paper is also on these two cases.

This entropy bound is indeed saturated generally in some occasions for infinitesimal perturbation of vacuum,
as later tested in \cite{Blanco:2013joa} the saturation of the bound (\ref{CBbound}) to the first order variation in the AdS/CFT framework. Also, Dvali recently showed that the saturation of universal entropy bounds is also related to the unitarity of scattering amplitudes \cite{Dvali:2020wqi}.  
 
In this paper, under the semi-static process to extract gravitational force by fixing local measurement of $\Delta \braket H\rightarrow m$ for nearby static observers,
we show the saturation of the entropy bound 
\bea
\Delta S= {\Delta \braket{H}\over T}\,,
\label{BCbound}
\eea
leads to an entropic gradient generally
\bea
\nabla_\mu S= \frac{m}{T_0}\nabla_\mu V\,,
\label{entgrad}
\eea
where $m$ is the mass of the test particle and $V=e^\phi$ is the redshift factor with respect to the general gravitational potential $\phi$,
while $T$ standing alternatively for the local measurement of the Unruh temperature or the Hawking temperature for static observers along with $T=T_0$ for $V=1$. It proves the necessity of external force for an entropic mechanism.

The covariant version of the external force (necessary to balance the gravitational force) is emerged directly, from the entropic force formula 
\bea
F_\mu=T\nabla_\mu S\,,
\label{entforce}
\eea 
rather than the gravitational force as the inertial force in Verlinde's original theory.

While this entropic force formula is no longer macroscopic effect after involving fine-grained entropy bound, it is just an approximation of a more general modular Hamiltonian approach we develop. Indeed, the true derivation of the inertial force $F_g$ as emergent gravity actually comes from utilizing the entanglement first law to get a work term
\bea
dW_g=-d\braket{O}_1\,,
\eea
where $O=K_1-K_0$ is the difference between the modular Hamiltonian of excited states and vacuum states. Here, we further prove when the bound is saturated, the resulted inertial force doesn't dependent on the detail of $O$. The variation of Casini-Bekenstein bound in such quasi-static processes will naturally reproducing Newton's Second Law in Rindler space and local gravitational force for Schwarzschild black hole.

After this, we reached the core topic is to find a holographic interpretation for gravitational attraction, since the saturation of this bound is a condition of holography. 
Noticing the connection between the saturation of the Casini-Bekenstein bound and the first law of black hole thermodynamics, we interpret the entropic gradient holographically as
\bea
\nabla_{\mu}S=\nabla_{\mu} \left(\frac{\delta A(\Sigma_{r_s})}{4G}\right)\,,
\eea
corresponding to the variation of horizon area $\delta A(\Sigma_{r_s})$ as extremal surface, rather than the variation $\delta A(\Sigma_{r})$ of the "holographic screen" at $r$,  which would otherwise provide too much holographic entropy. This holographic interpretation is covariant, and corresponds to the Bousso bound in \cite{Bousso:1999xy,Bousso:1999cb} (reviewed in \cite{Bousso:2002ju}).

\paragraph{Extremal Surface}
During dynamic processes, such as the black hole evaporation and matters free-fall towards the black hole, the Bousso bound  which is covariant associated with the extremal surfaces stays the same. 
In a holographic theory with the AdS/CFT correspondence \cite{Maldacena:1997re, Witten:1998qj}, it is the Ryu-Takayanaki surface (and covariant HRT surface) as well as its quantum versions that corresponds to the generalized entropy. 
We point out the new holographic interpretation of the entropic gradients will reflect on the variation of extremal surfaces in this framework.

\paragraph{Page Curve from the Entropic bound}
 While, processes with external influence viewed as heat flux into/out of reservoirs will vary extremal surfaces, as recent considered to evaporate AdS black holes for Black Hole Information Paradox in \cite{Almheiri:2018xdw,Penington:2019npb}.

We suggest at the same time the entropy change because the associated gravitational attraction is canceled. 
And we find a function 
\bea
S_{ext}=\frac{m_{rad}}{T_H}=8\pi G(M-m_{rad})m_{rad}\,,
\eea
possessing the expecting shape property of the Page Curve along with the local temperature increasing during the evaporation. 

\paragraph{Complexity Tendency}

Recently, Susskind argued in \cite{Susskind:2018tei} that gravitational attraction comes from the complexity tendency \cite{Susskind:2014rva} by proposing Size-Momentum Duality \cite{Susskind:2018tei} and claimed it is not compatible with an entropic mechanism that may be not able to explain the oscillation of free particles in pure AdS  \cite{Susskind:2019ddc} . 

Again the salvation is natural after our theory: these two kind of theories are in two considerations of processes and indeed they can co-exist after distinguishing situation difference.  We show a proper adapted Emergent Gravity theory to AdS may help understanding gravitational attraction in pure AdS, and more over, possibility is there to build a connection to transform between the entropic gradient and the operator growth, once we know the generic entropic gradient in spacetime and turn it into momentum-change, through virtual processes involving intermediate states.
\paragraph{Structure of the Content}

The structure of the paper is as follows. 

In Section \ref{EnT}, we set our stage by reviewing Casini-Bekenstein bound, for bipartite systems. Then we show how the entropic gradient raises and reproduces results matching GR.

In Section \ref{EmI}, we will further develop the techniques to derive inertial force utilizing the entanglement first law, which is more rigorous, and compare it with the derivation from the entropic force formula. Then we introduce our new holographic interpretation for the entropy change to explain gravitational force, noticing the connection between the upper entropy bound and the first law of black hole thermodynamics.

In Section \ref{SpI}, we show the implications from the understanding of gravitational force though our mechanism. We compare our results with Verlinde's original proposal. Then we move to the black hole information problem, to see the potential connection if it involves the same kinds entanglement responsible for gravitational attraction in our theory. Finally, we specify the occasional difference between Emergent Gravity and Complexity Tendency.

In the whole context, we adopt the Natural Unit $c=k=\hbar=1$ unless otherwise specified.

\section{Entanglement and Thermodynamics}
\label{EnT}

In this section, we set our stage on cases of bipartite systems, whose Hilbert space admits a tensor factorization $\H = \H_{A}\otimes \H_{\bar A} $. We consider relativistic Quantum Field Theories on a stationary geometry background with metric $ds^2=g_{\mu\nu}dx^\mu dx^\nu$. For them, such a decomposition is not arbitrary, according to the Reeh-Schlieder theorem. Then, we review the Casini-Bekenstein bound, a general result for any relativistic QFT that respects such decomposition. The modification of such bound requires changing the modular flow, which is supposed to be conserved during unitary transformation. We will show no change of local quantity $\Delta H$ is the specific condition that leads an entropic gradient which can reproduce Newton's 2nd Law and gravitational force in GR, as in the two cases we are familiar with definition of this bound, Rindler space and static black hole,  .

\subsection{Casini-Bekenstein Bound in Global Causal Wedges}

For any global state with density matrix $\rho=|\Psi\rangle\langle \Psi |$ in a general quantum system, the state confined in the subsystem $A$ (whose complement is $\bar{A}$) can be described by the reduced density matrix $\rho_A=\Tr_{\bar{A}} \rho$. We can always write the reduced density matrix as
\bea
\rho_A=\frac{e^{-K}}{\Tr{e^{-K}}}\,,
\eea
because it is positive defined and hermitian. $K$ is known as the modular Hamiltonian \cite{Haag:1992hx} of $\rho_A$. The entanglement entropy is defined as the von Neumann Entropy
\bea
S(\rho_A)=-\Tr\rho_A\log{\rho_A}\,.
\label{entS}
\eea

Let us consider the special cases in relativistic QFTs whose Hilbert space can be decomposed as a tensor product $\H=\H_{R}\bigotimes \H_{R^c}$, associated to spatial region $R$ (which has an algebra $\mathcal A(R)$ of local operators) and its complementary set $R^c$ lying on a Cauchy slice. By tracing over $\H_{R^c}$, we get the reduced density matrix
\bea
\rho_R=\text{Tr}_{R^c} \rho
\eea

Generally, such tensor decomposition in relativistic QFTs is not possible if $|\Psi\rangle$ is cyclic and separating \cite{Witten:2018lha}, according to the Reeh-Schlieder theorem. Special global causal wedges such as Rindler wedge in Minkowski spacetime are where the decomposition can take place. Therefore, we would rather set the thermodynamics in global causal wedge, rather than forming a local entropic mechanism by thermodynamics on "local Rindler horizon" as in \cite{Jacobson:1995ab}.

We denoted the causal domain of $R$ as $D(R)$. While choosing another spatial region $V'$ which shared the same causal domain $D(R')=D(R)$, the entanglement entropy 
stay the same 
\bea
S(\rho_{R'})= S(\rho_{R})
\eea
and it doesn't change under unitary transformations $U$
\bea
\rho_{R'}=U^\dagger \rho_R U\,,
\eea
Also, during time evolution, the unitary transformation doesn't change the entanglement entropy inside of the causal domain. 

Take the half space $R=\{t=0,x\geq0\}$ in Minkowski spacetime for example first, its causal domain is the Rindler space, called the right Rindler wedge. According to \cite{Bisognano:1976za}, the Minkowski vacuum state confined in the right Rindler wedge is a Gibbs state
\bea
\rho_R^0=\frac{e^{-{H/ T}}}{Z}\,,
\eea
and the modular Hamiltonian of the vacuum state is the boost generator $K=\frac{H_\eta}{T_U}$ , which is a local operator and generates a conserved modular flow. 
We can see this from the conserved charge $\int_{\Sigma}T_{\mu\nu}\chi^{\mu}d\Sigma^{\nu}$ associated with the Killing vector $\chi^\mu$, thus the expectation value of modular Hamiltonian generates the conserved flow from the local operator
\bea
 H=\int_{\Sigma} T_{\mu\nu}\chi^{\mu}d\Sigma^{\nu}\,.
\label{Hchi}
\eea
We use the expectation value \bea
\braket{H}_{\rho_R}=\Tr \rho_R H
\eea
 to replace the role of energy, for the state labeled by its density matrix $\rho_R$. Since $\chi^\mu$ is dependents on the trajectory labeled by $\xi=const$, this expectation value are also related by redshift factor $V$ to different observers.

Generally, the vacuum fluctuation will causes UV-divergence in $S(\rho_R^0)$. Energy and entropy subtracting the vacuum fluctuation defined in \cite{Marolf:2003sq,Marolf:2004et} are
\bea
\Delta\braket{H}=\Tr{\rho_R^1 H}-\Tr{\rho_R^0 H}\,,
\eea
and
\bea
\Delta S=S(\rho_R^1)-S(\rho_R^0)\,.
\eea

Now, let us review Casini's proof. The relative entropy is defined as
\bea
S(\rho|\sigma)=\Tr{\rho\log{\rho}}-\Tr{\rho\log{\sigma}}~.~~~
\eea
and from the positivity of the relative entropy
\bea
S(\rho_R^1|\rho_R^0)=\Delta\braket{K}-\Delta S\geq0\,.
\eea
Casini simply proved 
\bea
\Delta S\leq\Delta\braket{K}\,,
\eea
 which is
\bea
\Delta S\leq\Delta\braket{H}/T\,.
\label{SH/T}
\eea
when including thermalization. 

In the whole context, we always take the saturation of the entropy bound
\bea
\Delta S=\Delta\braket{K}\,,
\eea
or
\bea
\Delta S=\Delta\braket{H}/T\,.
\eea
Now we set up the stage and the definition of quantities.

\subsection{Where does the Entropic Gradient come from?}
\label{ThermoCondi}



This question directly links to the interesting query how one can realize gravitational force as a thermodynamic force.

Between two static observers with different trajectories label by $\lambda'$ and $\lambda$, the local measurement of the conserved quantity $\Delta \braket {H}$ and temperature $T$, both  depend on the redshift factor accordingly
\bea
H'/H=T'/T=V(\lambda)/V(\lambda')\,,
\eea
where the second equality is for the Tolman Law, while it is $K=H/T$ that stays the same.
But do remember the entanglement entropy is always the one in this Cauchy slice,  so we write
\bea
\Delta S'=\Delta S\,,
\eea
even if the entropy bound is not saturated.

Let us define
\bea
\delta\Delta \braket {H}=\Delta\braket {H'}-\Delta\braket {H}
\eea 
and
\bea
\delta \Delta S=\frac{\Delta\braket {H'}}{T'}-\frac{\Delta\braket {H}}{T}
\label{dDSm}
\eea
for infinitesimal variation.

The thermodynamics comes when one tries to extract gravity,
in processes under a special condition
\bea
\delta\Delta \braket {H}=0\,,
\label{H0cond}
\eea
which will cause the entropy bound change
\bea
\delta \Delta S=\delta\frac{\Delta\braket {H}}{T}=\frac{\Delta\braket {H}}{T_0}\delta V\,.
\label{dDSmV}
\eea
This condition reveals the origin where the entropic gradient comes into the story.

Or we consider what happens in the view of the same observer with $H$. Then temperature is fixed $T'=T$ but after the influence, the condition (\ref{H0cond}) is equivalent to $\Delta\braket{H}'=\Delta\braket{H}V$, so we will still get (\ref{dDSmV}).

The expectation value of $H$ is the integration
\bea
\braket{H}_{\rho_R}=\int_{\Sigma}\braket{T_{\mu\nu}}_{\rho_R}\chi^{\mu}d\Sigma^{\nu}
\eea
of the expectation value of local operator $T_{\mu\nu}$. Thus one test particle (we call it "box") as excited state localized at the position of one local observer can be made by centralizing/massing $\braket{T_{\mu\nu}}_{\rho_R}$ into small region. 


It is the external influence to overcome the redshift effect that brings in thermodynamics to form an entropic mechanism for gravity. The external influence then causes the heat flow $\delta Q=T\delta \Delta S$ into the causal wedge. It is easy to ignore that this process is not unitary, if  the progress changes the fine-grained entropy in the whole casual wedge. 



\subsection{Emergence of Newton's 2nd Law in Rindler Space}

In the coordinate $\{\eta,\xi\}$, the metric of Rindler space is
\bea
ds^2=e^{2a\xi}(-d{\eta^2}+d{\xi}^2)\,,
\label{RindlerI}
\eea
for the right Rindler wedge of the Minkowski spacetime.

Every orbit $\xi\equiv const$ corresponds to one of the different accelerating observers following a boost killing vector $\d_\eta$. Those accelerating orbits share the same Rindler horizon $H^\pm$ as well as the same causal development, which is the right Rindler wedge.

The redshift factor is
\bea
V(\xi)=\sqrt{-\chi_\mu \chi^\mu}=e^{a\xi}
\eea
where $\chi_\mu$ is the killing vector.

The surface gravity of the Killing horizon of the wedge is just
$\kappa=a$, so the Unruh temperature \cite{Unruh:1976db} is
\bea
T=T_U=\frac{a}{2\pi}\,,
\eea
where the parameter $a$ is also the acceleration of the observer following the orbit $\xi\equiv0$. 

From the proposed entropic gradient expression (\ref{entgrad}), we will get
\bea
\nabla_\mu S=
\frac{m}{T_U}\delta^{\xi}_{\mu}\d_\xi V(\xi)
=
\delta_{\mu}^{\xi}2\pi m e^{a\xi}\,,
\eea
and the entropic force formula (\ref{entforce}) produces
\bea
F_\mu=T_U\nabla_\mu S=\delta_{\mu}^{\xi} m a
 e^{a\xi}\,,
\eea
where the covariant $\delta_{\mu}^{\xi}$ shows the force is in the direction to switch the orbit towards the one with higher acceleration. 
So the external force $F=\sqrt{F_\mu F^\mu}$ is
\bea
F=ma\,,
\eea
which exactly agrees with Newton's 2nd Law.

\subsection{Emergence of Gravitational Force}

We set a stationary background of asymptotic flat Schwarzschild black hole with the metric
\bea
ds^2=-(1-\frac{2GM}{r})dt^2+{1\over(1-\frac{2GM}{r})}dr^2+{r^2}d\Omega^{2}\,,
\eea
in the global coordinate. We ignore the back-reaction from our test particle to the geometry.

The redshift factor is
\bea
V(r)=\sqrt{-\chi^\mu\chi_\mu}=\sqrt{-g_{00}}=\sqrt{1-\frac{2GM}{r}}\,,
\eea
the entropic gradient is
\bea
\nabla_\mu S &=&\frac{1}{T_H}\frac{GM}{r^2\sqrt{1-\frac{2GM}{r}}}\delta^r_\mu\,,
\label{Sgenr}
\eea
and the local measure the Hawking temperature for the static observer with $r\equiv const$ is
\bea
T=\frac{T_H}{V(r)}\,.
\eea
So the entropic force formula reproduces
\bea
F_\mu=T\nabla_\mu S=\frac{GMm}{r^2(1-\frac{2GM}{r})}\delta^r_\mu\,.
\eea
For the observer at infinity, the force amounts $F=V(r)\sqrt{F_\mu F^\mu}=\frac{GMm}{r^2}$.

Notice that it is directly covariant results calculated in General Relativity, see textbook \cite{Carroll:2004,Wald:1984}. These results agree with the local external force $\textbf{F}_{ex}$ calculated in General Relativity
\bea
\textbf{F}_{ex}=ma_\mu
\eea
with $a^\mu=U^\nu\nabla_\nu U^\mu$, for the static observer whose four-velocity $U^\mu$ is proportional to the time-translation Killing vector $\d_t$. 
\paragraph{Near-Horizon Limit is not generalizable}
We note that, to form a general entropic mechanism, the local Hawking temperature $T(r)=\frac{T_H}{V(r)}$ plays the ordinary role of the Unruh temperature $T_U$ in Entropic Gravity theories.  And our results are directly consistent with the gravitational force, not just in the near-horizon region. 

In the near-horizon limit, the black hole geometry approximates the Rindler space while the local Hawking temperature approximates the Unruh Temperature, that's why an entropic mechanism works directly in generic situations can be applied to the near-horizon region, not the other way around.



\section{The Emergence of Inertial Force}
\label{EmI}

In this section, we develop the entropic mechanism in detail to derive the inertial force from the entanglement first law. It is a specific technique to extract gravitational attraction through thermodynamics. Then we give our new holographic interpretation, after confirming that the saturation of the Casini-Bekenstein bound is closely related to the first law of black hole thermodynamics, providing exact amount of entanglement entropy necessary for generic situations.



In Newton's Mechanics, to maintain any object of mass $m$ relatively static to the accelerating/inertial frame with acceleration $a$,
we need to add on one external force
\bea
F=ma\,,
\eea
which is a reframed statement of the Newton's Second Law. While, from the point of view of one accelerating observer, the balance condition
\bea
F_i+F=0
\label{fifex}
\eea
should be satisfied for one effective force, $F_i$, which is the inertial force.

However, in General Relativity, 
 the free-falling trajectory is indeed geodesic with no acceleration. We choose the accelerating frame to be static, with the velocity $U^\mu$ proportional to the time-like Killing vector $\chi^\mu$. The acceleration $a^\mu=U^\nu\nabla_\nu U^\mu$ is for the static observer following an time-like killing vector, and then gravitational attraction becomes the inertial force
\bea
\textbf{F}_g=m g^\mu
\eea
where $g^\mu=-a^\mu$ is the gravitational acceleration, for the geodesic relative to that static observer.

To calculate the inertial force from thermodynamics, let us form a quasi-static process to move the object a little bit to the nearby trajectory, with the existence of external force satisfying the balance condition $\textbf{F}_{ex}+\textbf{F}_g=0$. Noted that, this process will not change the momentum
\bea
\frac{dp}{d\lambda}=0\,,
\label{dp0}
\eea
which is the major divergence from Susskind's situation for Complexity Tendency.


\paragraph{Modular Hamiltonian}
We use the expectation value of the modular Hamiltonian as "energy" in the spacetime thermodynamics. We already know the Killing vector $\chi^{\mu}$ is associated with a conserved charge
\bea
E_T=\int_{\Sigma}T_{\mu\nu}\chi^{\mu}d\Sigma^{\nu}\,,
\eea
In Rindler space, this leads to the boost generator
\bea
H_{\eta}=a\int_{x>0}d^{d-1}x \, x\, T_{00}\,,
\label{Heta}
\eea
for the Killing vector
\bea
\d_\eta=a(x\d_t+t\d_x)\,,
\eea
to the observer of acceleration $a$. And $K=H_\eta/T_U$ is the modular Hamiltonian of the vacuum state $\rho_R^0$. FOr example, the vacuum state for eternal black hole without radiation is Hartle-Hawking state \cite{Hartle:1976tp} 
\bea
\rho_{HH} \sim e^{-H/T_H}
\eea
where $H$ is the time-translation symmetry operator for the static geometry as (\ref{Hchi}) associated with the Killing vector $\d_t$ for the observer at infinity.

Now we would also clarify that thermodynamics for spacetime is always associated with the quantum expectation value $ \braket{H}$ along with the temperature $T_H$, neither classical Komar mass nor ADM mass. Macroscopic thermal temperature is probably irrelevant here. However the conserved quantum quantity $\Delta \braket{H}$ will promisingly approximate to Komar mass or ADM mass in the classical limit.

\subsection{External Work Term from The Entanglement First Law}

In the previous work \cite{An:2018hyt}, we derive certain thermodynamic equations to calculate the inertial force, noticing the difference between the thermodynamics first law and entanglement first law. Let us illustrate it in this subsection and then further develop it in the next subsection.

In the last section, we have set our stage on the causal wedge $D(R)$ associated to a special separation of Hilbert space $\H=\H_R\otimes \H_{R^c}$. The spatial region $R$ can be the half plane $x>0$ in Minkowski spacetime or the exterior region $r>r_s$ of two-sided Schwarzschild black hole. This stage allows us to form certain equations for thermodynamic quantities, by using the entanglement first law in the whole wedge.


The entanglement first law states that if $\rho_R(\lambda)$ of a state in the subsystem $V$ varying with one parameter $\lambda$, to the first order perturbation $d\lambda$ at $\lambda=\lambda_0$, we always have the following equation
\bea
\frac{dS(\rho_R)}{d\lambda } =\Tr \left(\frac{d\rho_R}{d\lambda} {K_R}\right)
\label{firstlawK}
\eea
or we can rewrite it as
\bea
dS=d\braket{K_R}
\eea
where $K_R=-\log{\rho_R(\lambda=\lambda_0)}$ is the modular Hamiltonian of the initial state. A detail proof can be find in \cite{VanRaamsdonk:2016exw}. As a consequence of (\ref{firstlawK}), we could take the parameters such as temperature $T$ in $K=H/T$ out of the derivative
\bea
TdS=d\braket H\,.
\eea
We note here there were some relevant papers about first-law-like relation for entanglement entropy. In \cite{Guo:2013aca}, the entanglement temperature was defined and the generalized entanglement first law relation in Gauss-bonet gravity and Love-lock gravity was studied. And in \cite{Lee:2010bg,Lee:2010fg}, entanglement entropy and a first-law-like relation was introduced to explain gravitational force from information erasing. While, they didn't involve modular Hamiltonian.
\paragraph{The work term}
Now we write the entanglement first law for the vacuum state $\rho_R^0=e^{-H/T}/\Tr e^{-H/T}$ as
\bea
T d S_0= d \braket {H}_0\,,
\label{entf0}
\eea
and for the excited state $\rho_R^1=e^{-K_1}/\Tr e^{-K_1}$ as
\bea
T d S_1= d \braket {H}_1+Td \braket {O}_1\,,
\label{entf1}
\eea
where we take the modular Hamiltonian $K_1$ of the following form 
\bea
K_1=H/T+O\,,
\eea
where the operator
\bea
O=K_1-K_0
\eea
is the difference between the modular hamiltonian of $\rho_R^1$ and $\rho_R^0$ .

Subtract (\ref{entf0}) from (\ref{entf1}), we get
\bea
Td\Delta S=d\Delta \braket{H}+ Td \braket {O}_1
\eea
Compare with the thermodynamic first law $dW+dQ=dE$, one can easily make the hypothesis that the work term is related to
\bea
dW_g=-Td \braket {O}_1
\eea
which is the difference between the modular hamiltonian of $\rho_R^0$ and $\rho_R^1$.

By considering the variation of the state in the existence of the external influence, we can extract the work term $dW_g$ we claim accounts for gravity. We noted the detail form of modular Hamiltonian for excited states are given in \cite{Balakrishnan:2020lbp}. It supports our hyposis of $K_1=K_0+O$, and $O$ only involves in local operators in the visible causal wedge. In another work, the modular Hamiltonian for holographic excited states is also discussed in \cite{Arias:2020qpg}. 

While so far, we haven't apply the condition (\ref{BCbound}) $\Delta S=\frac{\Delta\braket H}{T}$ yet. We will prove after applying this condition, the external work will not depend on the detailed form of $O$.

As a good example, we provided a simple scalar model with single-frequency mode in the previous paper \cite{An:2018hyt}, to show explicitly what each term involved in is and how they vary during the process.
Accidentally after applying the saturation of entropy bound, during the quasi-static thermodynamic process below, successfully $dW$ term turns into the correct expression for the inertial force as gravitational attraction, rather than external force. 


\paragraph{Local isoenergic process vs global isothermal process}

Let us now explain the thermodynamic progress first proposed in \cite{An:2018hyt} in detail. We will see it is either an isoenergic process or an isothermal process in the eyes of different observers.

In the Bekenstein Thought Experiment (see a review such as \cite{Bousso:2002ju}), Bekenstein considered a quasi-static progress to classical level (historically it was called Geroch progress), to lower a box towards the black hole with a long string very slowly till Planck-scale-near the horizon and finally to drop it into the black hole.

While, beyond this near-horizon region, it could be still a thermodynamic process. Semi-classically, we consider the Hawking/Unruh effect that thermalizes the "box" (we take as an excited particle state). 
Once the gravitational force is balanced by the external force. In order to form a thermodynamic process which changes the states, Alice varies the static trajectory $X(\lambda_0)$ a little bit to the nearby trajectory $X(\lambda)$. So the infinitesimal variation $d\lambda$ of the states is to the temperature
\bea
\frac{d}{d\lambda}=\frac{dT}{d\lambda}\d_T\,.
\eea

In this quasi-static process, it is the external force that maintains the local measurement of frequency $\omega$ of the box not varying
\bea
\omega=\omega_A
\eea
to the local observer (let us call her the proper observer Alice) moving along with the box, so the local measurement of the  energy $E=2\pi \omega$ also stays the same.

The proper observer Alice who follows the "box" will endure a temperature field with the parameter $\lambda$
\bea
T_A&=&\frac{T(\lambda_0)}{V(\lambda) }V(\lambda_0)=\frac{T_0}{V(\lambda) }\\
\omega_A &\equiv &\omega
\eea
where $T_0=T(\lambda_0)V(\lambda_0)$ is the reference temperature to $V(\lambda)=1$. 


Or the process is equivalent to the fixed observer Bob, who will see the temperature fixed, but frequency changed when Alice moving with the "box"
\bea
T_B &\equiv &T(\lambda_0)\\
\omega_B &=&\frac{\omega_A}{V(\lambda_0) }V(\lambda)\label{BobFre}
\eea
The derivation is with respect to the frequency $\omega$
\bea
\frac{d}{d\lambda}=\frac{d\omega}{d\lambda}\d_\omega
\eea
For Alice and Bob, the distribution varies in the same way during the process, since the distribution factor varies as
\bea
e^{-\frac{\omega N}{T}}\rightarrow e^{-\frac{\omega_0 NV(\lambda )}{T_0}}
\eea
for both observer, with $N$ is the particle number of this frequency mode. This agrees with the statement that state for any time-slice in the same Cauchy slice is the same.

However, Alice forgets to include the redshift factor to measure global energy, if she insisting on using the local Hamiltonian to measure energy
\bea
H_A= H|_{\lambda=\lambda_A}
\eea
as if Alice think she is in flat spacetime (to use her measurement of frequency for energy $E=\omega$). In Alice's eyes, objects following geodesic will get gravitational redshift, while the frequency of the "box" keeps the same. 

Since the temperature increases for Alice, the expectation value for the fixed frequency has changed. Thus the "energy" changes with the temperature of the state
\bea
\frac{d\braket {H_A}}{d\lambda}=\frac{dT}{d\lambda} \Tr\left( H_A\d_T\rho_R\right)=\frac{dT}{d\lambda}\d_T \left( \Tr H_A\rho_R\right)
\eea
since the frequency in the distribution and Hamiltonian operator is fixed.

\paragraph {Emergence of the inertial force}
In this part, we will combine the saturation condition (\ref{BCbound}) during the Temperature-changing process, to see if the inertial force emerges the same as the entropic force formula as we used the entropic gradient. The derivation is independent of the detail form of $\rho_R^1$.
Let us rewrite the entanglement first law of the vacuum state (labeled by 0) and the excited state (labeled by 1) as
\bea
T d S_0=d \braket{H_A}_0\,,\\
d W_g+Td S_1=d \braket{H_A}_1\,.
\label{heat}
\eea
Subtracting the vacuum fluctuation will simply lead to
\bea
d W_g&=&-Td \Delta S+d\Delta \braket{H_A}\nn\\
&=&-Td \frac{\Delta(\Tr{H_A \rho_R})}{T}+d\Delta\braket{H_A}\,,
\label{TdS+dH}
\eea
where we apply the saturation of the bound (\ref{BCbound}) for the second equality
\bea
-Td \Delta S=-Td \frac{\Delta(\Tr{H_A \rho_R})}{T}
\eea
and we should be cautious that
\bea
d\Delta(\Tr{H_A \rho_R})&=& \Delta\Tr\{(d H_A) \rho_R\}+ \Delta\Tr\{ H_Ad\rho_R\}\,,\\
d\Delta\braket{H_A}&=&\Delta\Tr\{H_Ad\rho_R\}\,,
\eea
where $d H_A=0$ vanishes since the frequency doesn't change during the process. An example for this is in the single-mode scalar model in \cite{An:2018hyt}, where we have
\bea
H_A&=&\omega \N\\
O=K_1-K&=&\log \N
\eea
where the number operator $\N$ counting the particle number of the single frequency $\omega$ mode, so we would say $dH_A=0$ during this frequency-fixed process.

We end up with the work term simplified to
\bea                                                                                                                                                                                                                                                                                                                                                                                                                                                                                                                                                                                                                                                             
dW_g=-T\times {\Delta\braket{H_A}}d\frac{1}{T}\,,
\label{dW-}
\eea
which doesn't depends on the detail form of the operator $O$. Then local temperature field 
$T=\frac{T_0}{V}$ for this temperature-changing process leads to the inertial force
\bea
\textbf{F}_g=-T\times \frac{\Delta\braket{H_A}}{T_0}\nabla_\mu V\,.
\label{dF-}
\eea
This  formula is exactly opposite to the external force formula (\ref{entforce}) with the entropic gradient (\ref{entgrad}). And for Bob at fixed position with fixed temperature, the result will be the same, but $\nabla_\mu V$ comes from $\Tr dH \rho_R$, since the \textbf{isoenergy process} for Alice is a \textbf{isothermal process} with frequency varying according to (\ref{BobFre}) for Bob.

Noticing the minus sigh in (\ref{dW-}) and (\ref{dF-}), the approach using the entanglement first law will reproduce the inertial force, while the entropic force formula together with the entropic gradient will reproduce the external force, as we expect.

\paragraph{Compare with the entropic force formula}

When $T_A$ is very low such that the distribution factor $e^{-\omega/T_A}\ll1$, $\Delta\braket{H_A}$ stays almost the same 
\bea
\Delta\braket{H_A}'\approx \Delta\braket{H_A}
\eea
during the frequency-fixed process.
And the entropy bound varies almost the same way as  (\ref{dDSmV}) in the fixed-energy process
\bea
d\Delta S\approx\frac{\Delta\braket {H}}{T_0}d V
\eea
Thus this process approximates to the energy-fixed process in Section \ref{ThermoCondi} in low temperature limit. So we will still get
\bea
dW_g\approx-Td\Delta S\,
\eea
which is in the opposite direction to the change of the entropy bound $\Delta S$. 



\subsection{Connection to the First Law of Black Hole Thermodynamics}
The saturation of Casini-Bekenstein bound is the maximal entanglement entropy in the causal domain associated with the definite amount of "energy" within. Here we show it is closely related to the  first law of black hole thermodynamics: the upper bound for "box" outside of a black hole is also the increase of the holographic entropy when the "box" merging into the black hole.

For a static observer at $r$, the modular Hamiltonian $H$ associated with the Killing vector $\d_t$ at $r$
and local measurement of the Hawking temperature comes from the Tolman's law
\bea
T=\frac{T_H}{V(r)}\,.
\eea

If we introduce the following replacement to the entropy bound (\ref{BCbound})
\bea
T&\rightarrow & \frac{T_H}{ V(r)}
\\ \Delta \braket{H}&\rightarrow& m
\eea
where $T_H=\frac{\kappa}{2\pi} $ is the Hawking temperature with the surface gravity $\kappa=\frac{1}{4GM}$ for the Schwarzschild black hole, the entropy bound (\ref{BCbound}) becomes
\bea
\Delta S={\Delta \braket{H}\over{T}}\rightarrow \frac{m
 V(r)}{T_H}\,,
\eea
where we can import the detail form of $T_H$ to get
\bea
\frac{m
 V(r)}{T_H}=2\pi \times 4GM m V(r)= \frac{4\times2\pi(2GM) (2G mV(r))}{4G}
\eea
Since we know the Schwarzschild radius is $r_s=2G M$, we can write
\bea
\frac{m V(r)}{T_H}={8\pi r_s (2G mV(r))\over4G}\,.
\label{mV/T}
\eea
This result by introducing the $T_H$ reminds us to compare with  the 1st law of black hole thermodynamic. 

We can also rewrite the bound in a first-law-like form
\bea
T_H\Delta S_{}
= m V(r)\,,
\eea
while the first law of black hole thermodynamics \cite{Bardeen:1973gs} states
\bea
T_H \delta S_{BH}
= \delta M
\eea
if the change of black hole mass $\delta M$ relates to the change of Bekenstein-Hawking entropy
\bea
S_{BH}=\frac{A}{4G}\,,
\eea
where the area of event horizon is $A=4\pi r_s^2$, with the Schwarzschild radius $r_s=2GM$. Thus we know $\delta r_s=2G\delta M$  and
\bea
\delta  S_{BH}=\frac{\delta A}{4G} ={8\pi r_s \delta r_s\over4G}\,.
\label{dSbh}
\eea

By comparing (\ref{mV/T}) and (\ref{dSbh}), we can relate the change of the Bekenstein-Hawking entropy and change of black hole mass as following
\bea
\delta S_{BH}&=&\Delta S\\
\delta M&=& m V(r)
\eea
to the entropy bound in the causal domain and local measurement of mass by red-shifting to infinity.

 At the same time, we know the perturbation of the conserved energy in asymptotic flat Schwarzschild spacetime, is equal to the amount of local measurement of mass $m$ by red-shifting to infinity: $\delta M=m V(r)$. 
  Geometrically, the Schwarzschild radius will increase by $\delta r_s=2GmV(r)$, when the black hole absorbs the "box" completely with the local mass $m$ measured by static observer at $r$.

In a summary, the introduction of the local Hawking temperature made the entropy bound in the casual wedge equal to the change of the Bekenstein-Hawking entropy when black hole mass increases by $mV(r)$. The connection
\bea
\Delta S= \frac{\delta A(r_s)}{4G}\,
\label{dDS}
\eea
and 
\bea
\Delta \braket H= m\,
\eea
is the foundation to build the new holographic interpretation for our entropic mechanism.


In \cite{Bhattacharya:2018wfr}, the entropic force formula together with the entropic gradient that origins from the variation of horizon area, is tested through calculating the back-reaction to the geometry. They confirm the entropic force proposal works in the near-horizon region, for a large  Schwarzschild black hole, a large electrically charged black hole and slowly rotating Kerr black hole. However, they find the original "holographic screen" proposal doesn't work in generic situations.

Next, we show our discovery of (\ref{dDS}) here is the key to a new holographic interpretation beyond the near-horizon region.


\subsection{New Holographic Interpretation}
We have find that the upper bound of entropy to the mass $m$ of the box in a black hole background, is equal to the variation of the new black hole if merged with the mass $m$. And it corresponds to the radius variation of the event horizon by $\delta r_s=2GMmV(r)$. The saturation of the Casini-Bekenstein bound along with the vanishing relative entropy is equivalent to a more general condition of holography, for the matter exterior of the black hole horizon.

We can rewrite the Bekenstein-Hawking entropy
\bea
S_{BH}=\frac{A(\Sigma_\text{hor})}{4G}
\eea
and the event horizon can be regarded as the minimal surface $\Sigma_\text{hor}$ for two-sided AdS black holes.


\paragraph{Quasi-Static to Covariant}

Once we withdraw external influence by setting $\textbf{F}_{ex}=0$, the heat flow stops: $\delta Q=T\delta S=0$.
If the quasi-static process stops at $r$ and the mass $m$ starting to free-fall towards the black hole, the entropy change of the new black hole will depend on the final position $r$
\bea
d (S_{bh}'-S_{bh})=d \frac{m}{T(r)}
\eea
The external force measured at infinity in General Relativity exactly matches with the expression
\bea
\textbf{F}_{ex}=T_H\nabla_\mu (S_{bh}'-S_{bh})
\eea 
From (\ref{dDS}), we can write local inertial force in a holographic expression
\bea
\textbf F_{g}\approx -\frac{T_H}{V(r)}\nabla_{\mu} (\frac{\delta A(\Sigma_{r_s})}{4G})\,.
\eea
We point out that, covariantly this interpretation corresponds to the variation of Bousso bound \cite{Bousso:1999xy,Bousso:1999cb}, since this is the same situation to collapse matters to form a new black hole.

The new thing here is that this shows any attempting generalization will fail, if using the area change $\delta A$ of the holographic screen at $r$. Otherwise, the original holographic interpretation from
\bea
\delta S=\frac{\delta A(\Sigma_r)}{4G}\,
\eea
gives too much entropy that the region interior of the holographic screen is already full of a black hole \cite{Bhattacharya:2018wfr}.  Our interpretation is the right answer to generic situations, and simply explains the reason: the original holographic screen approach only works in the near-horizon limit and can't be generalized directly.
\subsection{A Glimpse to Emergent Gravity in AdS}

Before further developing our theory in the AdS/CFT  framework in detail, which remains a future work beyond this paper, here we can still make prophecies about good properties that our entropic mechanism will have when adapted into this framework, benefiting from its well-established holography.

The major difference from asymptotic flat spacetime comes from that AdS/CFT would provide homologous CFT on the boundary dual to the quantum gravity in the bulk. Thus with the proper decomposition of the entire Hilbert space of CFT into $\H = \H_{B}\otimes \H_{\bar B} $, the entanglement entropy corresponds to a good geometric object in the bulk, knowing as the extremal surface. 

Besides, the vanishing of relative entropy was tested in \cite{Blanco:2013joa} to the first order perturbation, we would expect the entanglement entropy is a function of the modular flow ("energy") from the saturation of entropy bound.

Therefore,  we would expect a better description for our new holographic interpretation, corresponding to the variation of the extremal surface during the process to change energy in AdS. 

\paragraph{Extremal Surfaces}

When there is matter carrying entropy $S_{out}$ outside of a black hole, the generalized entropy
\bea
S_{gen}=S_{bh}+S_{out}
\eea
follows the Generalized Second Law (GSL)\cite{Bekenstein:1974ax}.

In AdS/CFT, it is the geometric subject called "extremal surface" $\gamma_B$ that corresponds to the $S_{gen}$
\bea
S_{gen}=\frac{A(\gamma_B)}{4G_N}+S_{bulk}(\gamma_B)
\eea
for a decomposition of boundary into subsystem $B$ and its complement $\bar B$.
The classical extremal surface for static geometry is the Ryu-Takayanagi surface \cite{Ryu:2006bv} which minimizes the bulk area $\gamma_B$, and the bulk contribution can be omitted since it is sub-leading. The HRT formula  \cite{Hubeny:2007xt} was proposed as a covariant version in classical level, while in quantum level, FLM was proposed in \cite{Faulkner:2013ana} and then the Quantum Extremal Surface \cite{Engelhardt:2014gca} with an extra maximin procedure.  For a two-sided AdS black hole, the horizon can be regarded as the extremal surface for the entanglement entropy between two copies of CFT.



During the evaporation of AdS black holes, covariant versions of extremal surfaces don't vary, neither in classical nor in quantum level. This is equivalent to that the entropy bound stays the same in the covariant situation when test particles freely fall toward the black hole as a unitary process. 

However, when extracting gravitational force in the bulk, we would expect that the generalized entropy changes, as well as the extremal surface associated with it. So we may  again use the entanglement entropy  for the decomposition $\H = \H_{B}\otimes \H_{\bar B} $ of the boundary CFTs,  to interpret inertial force thermodynamically.


Besides, our entropic mechanism may also work to explain the gravitational force in pure AdS as the saturation of Casini-Bekenstein bound is tested perturbatively in \cite{Blanco:2013joa} , using $\Delta \braket {K_B}$ for the Casini-Bekenstein bound, since there is no temperature.

In all,  in the AdS/CFT framework, the role of surface $\Sigma_{r_s}$ should be taken by the extremal surface, and a similar entropic gradient will reflect on the variation of the extremal surface.



\section{Further discussion: Spacetime Information}
\label{SpI}

So far, we have seen that the thermodynamic force we derived matches with inertial force as a consequence of the entanglement first law under certain conditions. Before moving further, we list major differences from thermal mechanics below: 
\begin{enumerate}
  \item Our mechanism for gravity relies on the entanglement entropy which is the fine-grained entropy and doesn't miss any detail of the state, while thermal entropy is coarse-grained.
  \item The Unruh/Hawking temperature is an observer dependent effect, and origins from the Bogolubov transformation. To formulate equations as (\ref{entf1}), our mechanism requires the modular Hamiltonian $K_1$ for any excited state thermalized in the following form
\bea
K_1=H/T+O\,
\eea
\end{enumerate}
for a good formation of thermodynamic equations. That's probably because the mechanism works for the state generated in the vacuum sector, such as adding one particle in the right Rindler wedge. Afterwards, this rigorous formulation can be adapted to more situations such as AdS/CFT.

To Interpret attraction as entropic force requires several properties of excited states and only under occasions to detect it using external influence. In this section, we will discuss how our entropic mechanism can help understand some issues about spacetime information.

\subsection{Return to Compare with Verlinde's Original Proposal }

First of all, the most important difference of the entropic gradient to generic situations is indeed it is in the opposite to direction of the original proposed one in \cite{Verlinde:2010hp}. 

By defining the generalized gravitational potential
$\phi\equiv \frac{1}{2}\log\{-\chi_\mu\chi^\mu\}$ and writing redshift factor as $
V=e^\phi=\sqrt{-g_{tt}}$, we can rewrite our results (\ref{entgrad}) as
\bea
\nabla_\mu S&=&\frac{1}{T_H}me^\phi\nabla_\mu\phi\,.
\label{nSphi}
\eea
with local temperature
\bea
T&=&T_He^{-\phi}\,.
\label{Tphi}
\eea
Generally $V=\sqrt{1-\frac{r_s}{r}}$ for Schwarzschild solutions, the entropic variation decreases with $r^{-2}$
\bea
\delta S=\partial_rSdr=8\pi mGM\partial_r{\sqrt{1-\frac{2GM}{r}}}dr= -\frac{8\pi m G^2M^2}{r^2}\delta x\,,
\label{dSdx}
\eea
where $\delta x=-\sqrt{g_{rr}}dr=-V^{-1/2}dr $ is the proper distance in the direction pointing towards the black hole.

In the near-horizon region $r\rightarrow r_s=2GM$, (\ref{dSdx}) becomes
\bea
\delta S|_{r\rightarrow r_s }=-2\pi m\delta x
\eea
And if we restore all the dimensional constants $T_H=\frac{\hbar c^3}{8\pi k_B GM}$, $r_s=\frac{2GM}{c^2}$ and $m\rightarrow mc^2$ from the very beginning, we get
\bea
\delta S&=&-\frac{8\pi k_B mG^2M^2}{\hbar c^3 r^2} \delta x,\\
\delta S|_{r\rightarrow r_s}&=&-\frac{2\pi k_B mc}{\hbar }\delta x.
\eea
Surely in this limit $\delta S$ approximates to Verlinde's proposal of entropic gradient in \cite{Verlinde:2010hp}
\bea
\Delta S= 2\pi k_B \frac{mc}{\hbar}\Delta x
\label{DSDx}
\eea
with an opposite sign (from here till the end of this subsection, we recover the unit from Natural Unit and use the original symbols).

Now let's see why our result has an opposite sign. In short, here the formula $F=T\nabla S$ gives external force rather than the gravitational force in Verlinde's proposal, so the entropy gradient has an opposite sign. However, this opposite sign shows the fundamental different consideration between coarse-grained entropy and fine-grained entropy as following statement:
\begin{itemize}
  \item Entropy variation in (\ref{DSDx}) originated from displacement of $m$ into the black hole by a distance $\Delta x=\frac{\hbar}{mc}$ far away from the horizon. Thus it is conjectured positive: the coarse-grained entropy of black hole increases after absorbing $m$. 

\item When we consider the combination system of a black hole and the test particle it attracts,  the fine-grained entropy doesn't change if being adiabatic. The Casini-Bekenstein bound of entanglement entropy changes, only when the external influence that cancels out the gravitational red-shift effect, changes the energy during quasi-statistic processes.
\end{itemize}

In all, the opposite sigh reflect the different direction between coarse-grained entropy increasing tendency and manipulating changing fine-grained entropy. We noted here though \cite{Plastino:2018krc} agreed with the  $r^{-2}$ behavior of gravitation force in Newtonian limit, its direction is still the same with coarse-grained entropy increasing direction. The situation is similar to the difference between a free-releasing adiabatic piston versus a reversible heat engine.

At the same time, since generally the entropic gradient has $r^{-2}$ dependence on the radial coordinate $r$, it denies Verlinde's original generalization of the entropic gradient in \cite{Verlinde:2010hp}
\bea
\nabla_a S= -2\pi \frac{m}{\hbar}N_a\,,
\label{nSN}
\eea
along with generation of the temperature
\bea
T=\frac{\hbar}{2\pi}e^\phi N^b\nabla_b\phi\,,
\label{Tephi}
\eea
where $N_a$ is the unit vector orthogonal to the screen. As we argued at the end of Section \ref{EnT}, the near-horizon-region limit can not be directly generalized to beyond. On the contrary, the generic result approximates to the near-horizon-region result in the limit. 

In parallel, we would argue that the entropic gradient along with the local Hawking temperature being an alternative description of potential gradient $\nabla_\mu \phi$, as an alternative description of the geometry in spacetime.

\paragraph{The Principle of Equipartition}

We find equipartition relation
\bea
\frac{1}{2}nk_BT=E
\eea
is always satisfied
by
\bea
\frac{1}{2}Nk_BT_H=Mc^2
\eea
with the degree of freedom $n$ characterized by $N$ "bits" on the "holographic screen" 
\bea
N=\frac{4\pi r_s^2c^3}{G\hbar}\,,
\eea
In the original thought, temperature $T$ was taken as Unruh Temperature $T_U$ in non-relative case and $N$ was hypothesized to increase with $r^2$, while our result shows in general it is local Hawking temperature $T_H$ and $N$ stays the same. So for the density of "bit" per Area on the "holographic screen" at $r$ decreases
\bea
N/A=\frac{ r_s^2c^3}{G\hbar}r^{-2} 
\eea
Besides, from the integral on the surface $S$
\bea
M=\frac{1}{2}\int_S T dN
\eea
we can also get the natural generalization of Gauss's Law (for asymptotic flat Schwarzschild solution)
\bea
M=\frac{1}{4\pi G}\int_S e^{\phi}\nabla \phi \cdot dA\,,
\label{GaussL}
\eea
thus our mechanism can be used to derive Einstein Equation more strictly.

\paragraph{Derive Einstein Equation}
Our approach is parallel to Entropic Gravity theories in the sense to derive Einstein Equation from thermodynamics, but it makes the derivation more reliable beyond near-horizon region.

In 1995, Jacobson \cite{Jacobson:1995ab} used Clausius Law $\delta Q=T_U\delta S$ and holographic entropy $\delta S\sim \delta A$, to derive Einstein Equation from thermodynamics for the case of null screens.
With the similar reasoning borrowed from Jacobson, Verlinde used the natural generalization of Gauss's Law (\ref{GaussL}) from (\ref{nSN}) and (\ref{Tephi}), which is not valid beyond near-horizon region, to show the Einstein Equation can be derived on the time-like screens. Thus this derivation of Einstein Equation is also only valid in the near-horizon region.

After we show (\ref{GaussL}) indeed comes from our specific entropic mechanism with the corrected generation (\ref{nSphi}) and (\ref{Tphi}) to generic situations, we confirm the Einstein Equation can be derived on the time-like screens beyond near-horizon region.

To the question \textit{What is the Entropy in Entropic Gravity?}\cite{Carroll:2016lku}, we would answer that it is Casini-Bekenstein bound, which bounds entanglement entropy associated with Hawking temperature, that accounts for Entropic Gravity.
Gravitational effect shouldn't rely on thermal entropy associated with macroscopic temperature.

\subsection{Black Hole Information Problem and Extremal Surface}
Extremal Surfaces for the covariant entropy bounds don't vary during the evaporation of the asymptotic AdS black hole, neither classically nor in quantum level.
That is because the boundary serves as reflecting boundary conditions.

Recently, Almheiri etc \cite{Almheiri:2018xdw} imagined a process of extracting Hawking radiation and then throwing it back to the other-side for a two-sided AdS black hole, through absorbing boundary conditions. Also the entanglement wedge reconstruction \cite{Czech:2012bh,Wall:2012uf,Headrick:2014cta,Jafferis:2015del,Dong:2016eik,Cotler:2017erl} was studied to understand how the ER=EPR \cite{Maldacena:2013xja} proposal work for the dynamic two-sided black hole.

Following this breakthrough, a new approach \cite{Almheiri:2019psf,Penington:2019npb,Rocha:2008fe} considered absorbing boundary conditions that couples an exterior auxiliary reservoir to a holographic CFT dual to an evaporating one-sided AdS black hole.
 They argued from the variation of the quantum extremal surface \cite{Engelhardt:2014gca} after the Page time to claim that the black hole interior can be reconstructed from the Hawking radiation, based on the entanglement wedge construction.

After reaching this absorbing boundary, the radiation then gets into the reservoir. The area of the extremal surface decreases along with the area of the event horizon during the evaporation.

Here we only point out absorbing boundary conditions play the same role as the external influence in our approach of Emergent Gravity, by extracting the energy into the exterior reservoir and thus change the entanglement entropy.
When the radiation is extracted, the amount of total energy and the area of the event horizon also change, as happens in the quasi-static process in our context. 

Once we are able to show it is the same entanglement nature that corresponds to both the Hawking radiation and gravitational attraction, which is beyond this paper, we may tell how information is carried by Hawking radiation from gravitational attraction.
Before that, we propose a candidate of Page Curve for thermal states with the increasing temperature, which could be either on the right way to understand black hole information paradox through attraction, or just a coincident calculation.

\paragraph{Page Curve for Asymptotic Flat Black Holes}

Our entropic mechanism seems to favor initial entanglement between test particles and the black hole attracting them. We notice that attraction should rely on the Casini-Bekenstein bound as the maximal amount of entanglement entropy that it can achieve.
This kind of entanglement may be not different between the entanglement of the black hole and the Hawking radiation.

Therefore, our entropic mechanism can provide the bridge to the feature of such kind of entanglement for black hole information paradox from the feature of gravitational force. What we clarified is when observer trying to realize gravitational attraction in space by external influence, the amount of such entanglement also changes. What we clarified is when observer trying to realize gravitational attraction in space by external influence, the amount of entanglement also changes.
The black hole radiation should carry information to not violate unitarity.


We suggest one way out for evaporating asymptotic flat Schwarzschild black holes is to use the entropy bound $\Delta S$ related to the Hawking temperature to characterize the entropy of the radiation, rather than regard the radiation as thermal gas.

By cutting the space into layers,
we selected one orbit $r=r_{A}$ for the static observer Alice close to the horizon. After radiating Hawking radiation of mass $m_{rad}$ passing this observer, it is equivalent to extract this part of energy into a heat reservoir, and the mass remains inside $r<r_{A}$ is $M-m_{rad}$. We know $m_{rad}$ is attracted by $M-m_{rad}$. By distinguishing the entanglement entropy from whose temperature, we proposed 
\bea
S_{ext}=\frac{m_{rad}}{T_H}=8\pi G(M-m_{rad})m_{rad}\,,
\eea
as a version of Page Curve. To Alice's view, the maximal entropy of radiation is entangled with the remaining black hole $M-m_{rad}$, thermalized with increasing $T_H=\frac{1}{8\pi G (M-m_{rad})}$. 
When the evaporation is over, Alice finds her in the flat spacetime with $T=0$ to infer that the total Hawking radiation become a pure state. 

While for anther observer Bob, the entropy may keep increasing. Choosing another orbit  $r=r_{B}\gg r_s$ for the static observer Bob,  the temperature almost doesn't decrease
 for there is some Hawking radiation in between, to maintain the gravitational attraction is still from the mass close to $M$. If $r_B=\infty$, that evaporation is almost not in causal access to him, so he won't able to decode information in the Hawking radiation.


As \cite{Akers:2019nfi} argued that Page Curve appears in some models while disappears in others, we suggest there surely will be Page Curve with the information carried by Hawking radiation. But to some other observer, this effect may not be observable.  

\subsection{Entropic Mechanism vs Complexity Tendency}
Here we distinguish the occasional difference for the entropic mechanism and Complexity Tendency, so they can co-exist with each other.


Entropic Gravity was proposed as a parallel between general relativities and thermodynamics, while Susskind came up with the idea that \textit{Things fall because there is a tendency toward complexity} \cite{Susskind:2014rva}, and proposed there is also a connection between gravity and complexity.
Then he conjectured Size-Momentum duality in \cite{Susskind:2018tei}. This conjecture was tested in SYK model during the falling process towards a charged black hole \cite{Brown:2018kvn}, in consideration of scrambling. And one detailed relation was recently given in \cite{Lin:2019qwu}
\bea
\frac{d\mathcal{C}}{dt}\sim p
\eea
where $\mathcal C$ stands for operator size and $p$ for momentum.

Further more, \cite{Susskind:2019ddc} claimed that there are limitations of Verlinde's entropic mechanism to explain gravitational falling in pure AdS, and doubted whether an entropic mechanism can explain the gravitational pull-to-the-center in cold empty AdS, or to a conventional zero temperature massive body in its (non-degenerate) ground state.
Recently \cite{Lin:2019kpf} showed how to do calculation specifically for pure AdS in the complexity picture, when there is no black hole horizon.

We again would say that two theories work in different processes: there is no heat flow $\delta Q=T\delta S=0$ for the free-falling process. The entropic explanation only works in quasi-static processes where external influence brings entropy changes. The major difference is that the momentum stays the same
\bea
\frac{dp}{d\lambda}=0
\eea
during the quasi-static process (or more accurate, the average momentum value stays the same). What's more, in the end of last section, we argue that our entropic mechanism can also work in AdS/CFT, even for gravitational force in pure AdS. 

Even more, through virtual quasi-static processes, we may be able to finally connect those two descriptions. The possible treatment may rely on chaining one more quasi-static process to give back the extracted gravity part $dW_g$ into momentum part $dW_p$ following fixed-frequency process in a time $dt$ necessary to moving $dr$, with no net effect in total
\bea
dW_g+dW_p=0\\
d\Delta S=0
\eea 
In this way, the entropic gradient transforms into complexity (operator growth) while gravitational potential energy transforms into momentum through virtual intermedia states in the entropic mechanism.

\section{Summary}
We build a more concrete entropic mechanism of the emergent gravity theory to explain gravitational attraction. It somehow differs from the original Entropic Force conjecture.


The entropic mechanism works under two major conditions:
\begin{itemize}
  \item This entropic mechanism appears under certain progresses. It requires external influence that causes the heat flow $\delta Q=T\delta\Delta S$ into the causal domain, thus varying the entropy bound. Under specific thermodynamics processes, gravity can be extracted and entropic gradient occurs.
  \item The saturation of the entropy bound turns the result matching with that of GR. Fine-grained entropy is thus introduced to explain the gravitational attraction. And this condition leads implications of spacetime information.
 \end{itemize}

The first condition, is consistent with the spirit of the Principle of Equivalence. Only when we trying to detect gravity though interfering, can we feel the existence of gravitational force. Now, it requires information change after considering the thermalization of the Hawking temperature. Otherwise, we admit that there will be no $\delta Q$ in unitary processes such as free-falling.

Figuring out the problem when will thermodynamic process arise allows us to distinguish the different occasions between Emergent Gravity and Susskind's Complexity Tendency: the latter doesn't vary the entanglement entropy in the whole causal wedge. So they can co-exist with each other, and we may find a way to connect the entropic gradient we calculated to operator growth through transforming.

For the second condition, the replacement of thermal entropy distinguish Emergent Gravity from macroscopic thermal mechanics.
This setup allows us to move the stage to quantum systems, and utilize the entanglement first law to form precise thermodynamic equations viewed by local static observers.

The Casini-Bekenstein bound supplies one simple relation between upper entropy bound and energy. After we point out it corresponds to the area variation of the horizon as the extremal surface, it leads to a simple holographic interpretation of the entropic gradient to explain gravitational attraction.

On the other hand, gravitational attraction is linked to the spacetime information. The entropic gradient can be regarded as an alternative characterization of geometry.
And also, it is possible that gravitational attraction may come from the same physics of ER=EPR. At least, we are talking about the same fine-grained entropy to explain gravitational attraction and to understand how Hawking radiation carries information. Then the entropic mechanism may help to understand black hole information paradox. 

These setups modify the way we think about Emergent Gravity theories to explain gravitational force. 
Afterwards, this could be a   rigorous formulation adapted to generic situations. To approve the mechanism in detail, we would expect it also works in AdS/CFT and can be verified in this better holographic frame. 
The mechanism can also serve for the bridge to apply results in \textbf{It's From Qubits} to cosmology through the observation of gravitational force. Above all, we have  solved/realized basics for Emergent Gravity to explain gravitational force in this paper.

\section{Acknowledge}
This paper started from a question.  Near the Christmas of 2017, YA asked Erik Verlinde a question: "What's the information meaning of distance?" However, till now we are still far from answering it.

We would like to thank many people for unconditional help and contribution to the paper, which is the positive side of doing research on physics.
The list is as following: Jan de Boer, Janet Ling-Yan Hung, Bartek Czech, Yu-Sen An, Shan-Ming Ruan, Zhenbin Yang, Ellis Ye Yuan, Huajia Wang, Gui Pimentel, Shira Chapman, Ben Freivogel, Adolfo Toloza, Bo Feng, Erik Verlinde for useful discussion and feedbacks.
And we do encourage people around world to cooperate together against Coronavirus without prejudice, as people do in science.

\bibliographystyle{JHEP}
\bibliography{Blackholereview.bib}

\end{document}